\documentclass[runningheads]{llncs}
\usepackage{graphicx}
\usepackage{booktabs}
\usepackage{multirow}
\usepackage[inline]{enumitem}
\usepackage{mathtools}
\usepackage{bbold}
\usepackage{floatrow}
\usepackage{adjustbox}
\usepackage{hhline}
\usepackage{bm}
\usepackage{xcolor}
\usepackage{wrapfig}
\usepackage{lipsum}
\usepackage{hyperref}
\DeclarePairedDelimiter\ceil{\lceil}{\rceil}
\DeclarePairedDelimiter\floor{\lfloor}{\rfloor}

\begin{document}
\title{Recommendation of Compatible Outfits Conditioned on Style\thanks{Supported by Flipkart Internet Pvt. Ltd.}}
%
%
\author{Debopriyo Banerjee \inst{1,2}\textsection\orcidID{0000-0001-9773-776X} \and
Lucky Dhakad \inst{2}\textsection\orcidID{0000-0003-3808-9015} \and 
Harsh Maheshwari \inst{2}\textsection\orcidID{0000--0002-9568-1093} \and
Muthusamy Chelliah\inst{2} \and
Niloy Ganguly\inst{1,3}\orcidID{0000-0002-3967-186X} \and
Arnab Bhattacharya\inst{2}\orcidID{0000-0003-2828-7691}}
\authorrunning{Banerjee et al.}
%
\institute{IIT Kharagpur, Kharagpur, India\\
\email{debopriyo@iitkgp.ac.in} \and
Flipkart Internet Pvt. Ltd., Bangalore, India. \and
Leibniz University Hannover, Hannover, Germany
}
\maketitle              
\begingroup\renewcommand\thefootnote{\textsection}
\footnotetext{These authors contributed equally to this work}
\endgroup
\begin{abstract}
Recommendation in the fashion domain has seen a recent surge in research in various areas, for example, shop-the-look, context-aware outfit creation, personalizing outfit creation, etc. The majority of state of the art approaches in the domain of outfit recommendation pursue to improve compatibility among items so as to produce high quality outfits. Some recent works have realized that \emph{style} is an important factor in fashion and have incorporated it in compatibility learning and outfit generation. These methods often depend on the availability of fine-grained product categories or the presence of rich item attributes (e.g., long-skirt, mini-skirt, etc.). In this work, we aim to generate outfits conditional on styles or themes as one would dress in real life, operating under the practical assumption that each item is mapped to a high level category as driven by the taxonomy of an online portal, like outdoor, formal etc and an image. We use a novel style encoder network that renders outfit styles in a smooth latent space. We present an extensive analysis of different aspects of our method and demonstrate its superiority over existing state of the art baselines through rigorous experiments.

\keywords{complete the look \and neural networks \and outfit compatibility \and style.}
\end{abstract}
\section{Introduction}\label{sec:intro}
Recommendation of outfits having compatible fashion items is a well studied research topic in the fashion domain \cite{bettaney2020fashion,chen2021tops,li2020bootstrapping,Li:2020:CompositionalVisualCoherence,Lin:2020:AmazonFOCIR,revanur21semi,yang2020learning}. Recent research in this regard explores graph neural networks (GNN) to connect users, items and outfits \cite{cui2019dressing,li20hier,liu2020learning,Wang:2021:GAttnNetVSE,yang20learning,Zhan:2021:A3-FKG} based on historical purchases as well as personalization \cite{chen2019pog,chen19pers,jaradat2020outfit2vec,landia2021personalised,Lin:2020:OutfitNet,lu2019learning,Lu:2021:CVPR} and explainability \cite{dong2020fashion,han2019prototype,Lin:2019,yang2019interpretable}. An apparent shortcoming of the current research on compatibility learning is the complete disregard for the explicit style associated with an outfit. However in real life, a person, say a user on an e-commerce platform, would typically have an explicit style in mind while choosing items for an outfit. The main objective of this paper is to learn compatibility between items given a specific style which in turn helps to generate style-specific outfits.

\begin{figure}[h]
    \centering
	\includegraphics[scale=0.12]{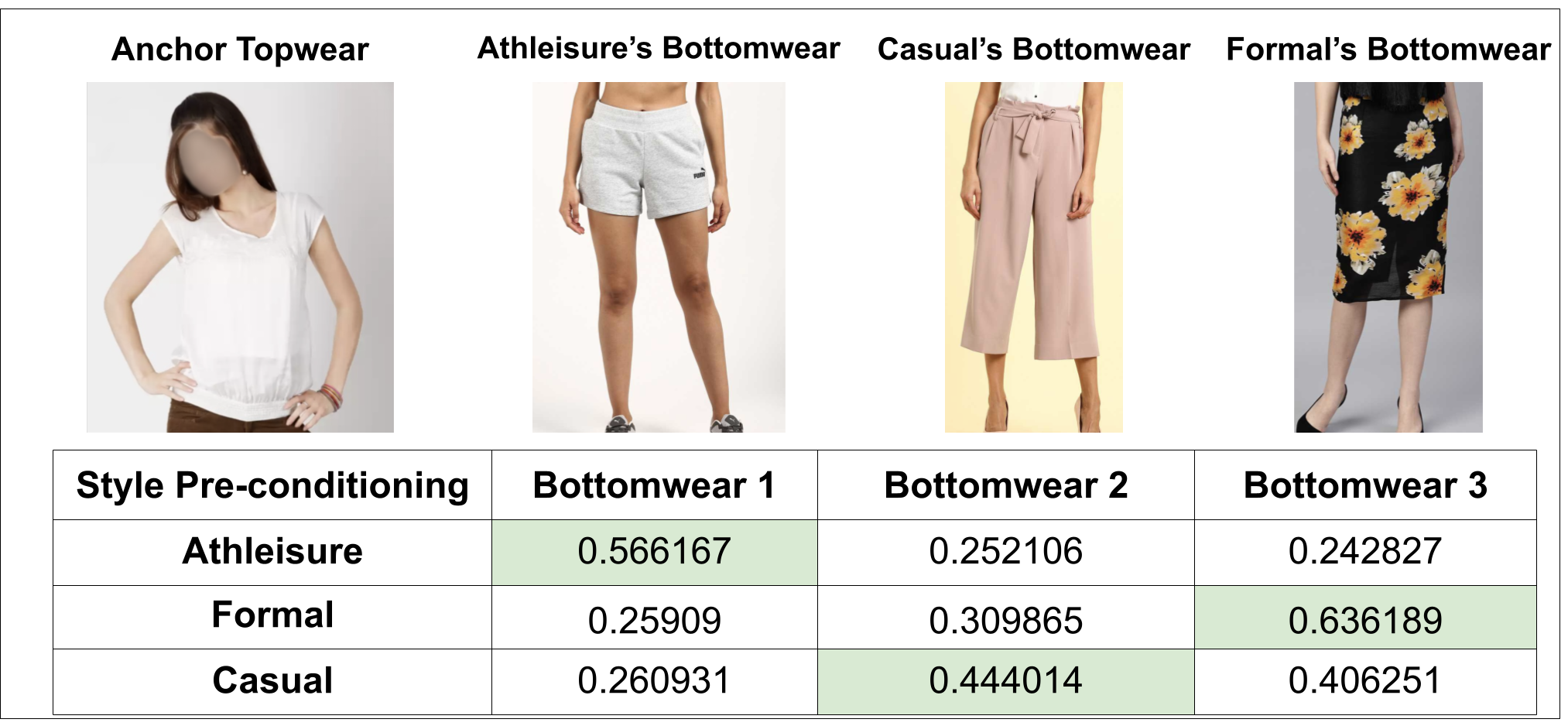}
    \caption{\small{Given a top-wear liked by a user, a style-guided method is able to create outfits conditional on various styles (\emph{athleisure}, \emph{formal} and \emph{casual}) while a style-independent compatibility model will typically generate outfits from dominant style. The values indicate the style-conditional compatibility scores for each item. Note that for a given style, the bottom-wear corresponding to that style gets the highest score.}}
    \label{fig:overview}
    \vspace{-2mm}
\end{figure}

We illustrate the importance of style-guided outfit generation through an example figure. Three sets of outfits are shown in Figure~\ref{fig:overview} with a white top-wear, an item that a user likes but is doubtful about making the final purchase (the reader is requested to ignore the values at the bottom of the figure for the time being). The platform may have the capability to showcase or to provide the user an option of generating outfits specific to various styles (this example showcases \emph{Athleisure}, \emph{Formal} and \emph{Casual}). Given this setup, a style-guided algorithm has two advantages: (a) it can generate compatible outfits from different styles, hence, providing the choice to the user, and (b) it will not generate an outfit which may be otherwise compatible but not in accordance with the desired style. The concept of jointly modelling for explicit style and compatibility is lacking in the area of fashion recommendation and current research have mostly treated them in separate silos. Having said this, one should be mindful of the fact that a style-independent compatibility algorithm followed by a style classification method, say Style2Vec \cite{Lee:2017:Style2Vec}, can allocate outfits to their relevant styles post the generation step. Thus in principle it is possible to combine existing work to generate the outfits in Figure~\ref{fig:overview}. It is however easy to see that such a technique is not efficient, since a large set of outfits need to be generated of which only a subset will be relevant to a particular style.
\vspace{-2mm}

\vspace{-6mm}

In recent times there have been some attempts at connecting style and outfit recommendation. Kuhn et al.\ \cite{kuhn:2019:Outfittery} does not consider the presence of explicit styles and rather learn compatibility while inferring the presence of latent style associated with each item. Jeon et al.\ \cite{Jeon:2021:FANCY} use extracted fashion attributes of full-body outfit images for modelling style classification, ignoring compatibility learning in the process. Learning outfit level theme or style from item descriptions, done by Li et al.\cite{li:2019:coherent} is a weak approach and fails when the descriptions do not exhaustively cover different styles. Singhal et al. \cite{Singhal:2020:VCP} models style between item pairs using an autoencoder, thus treating style as an implicit idea. A common deficiency in all of these works is the ability to generate style guided outfits.  Theme Matters~\cite{Lai:2020:ThemeMatters}, authored by Lai et al. is an archived work which comes closest to our model. It proposes a supervised approach that applies theme-aware attention to item pairs having fine-grained category tags (e.g., long-skirt, mini-skirt, etc.). The main handicap of their approach is that the size of the model increases exponentially with the number of fine-grained categories which was validated by our experiments.

We propose a {\bf Style-Attention-based Compatible Outfit Recommendation} (SATCORec) framework that uses high-level categories like top-wear, bottom-wear  etc. (general e-commerce taxonomy) and explicit outfit-level style information (\emph{formal}, \emph{casual}, \emph{sporty} etc) to learn compatibility among items in an outfit. It consists of two components, namely a Style-Compatibility-Attention Network (SCA Net) \cite{Lin:2020:AmazonFOCIR} and a novel Style Encoder Network (SE-Net). SE-Net considers an outfit to be a \emph{set} of items and makes use of the Set Transformer \cite{Lee:2019:SetTransformer} architecture to model a style specific distribution for each outfit. We believe that we are the first to adopt the set transformer, which is state-of-the-art technique to model data points that have the properties of a set, in a framework to project an outfit into a latent style space. Several variations of extracting a style representation from the learnt distribution have been investigated. We make use of this representation to estimate style-specific subspace attention within SCA Net which helps to learn compatibility conditional on style. Finally, we use the beam search approach \cite{Bettaney:2021:BeamSearch} to generate outfits based on a parent item, a template and a style.

We have created an in-house dataset of size approx.\ 100k corresponding to women's western wear outfits, taking items from an e-commerce portal. 
Various experiments have been performed on this data, comparing compatibility and style-specific metrics between baseline methods and SATCORec. Our method has been found to excel in compatibility learning, even when outfits are generated conditional on style. Most importantly, SATCORec is seen to outperform all the baselines in style metrics by a large margin.

\vspace{-4mm}
\section{Methodology}\label{methodology}
SATCORec is a deep learning model, developed to learn the compatibility between lifestyle items present within an outfit, contingent on the style to which the outfit belongs. The model first infers the style of the outfit which is subsequently used to learn compatibility between items within it.

\begin{figure}[ht]
    \centering
    \includegraphics[width=\linewidth]{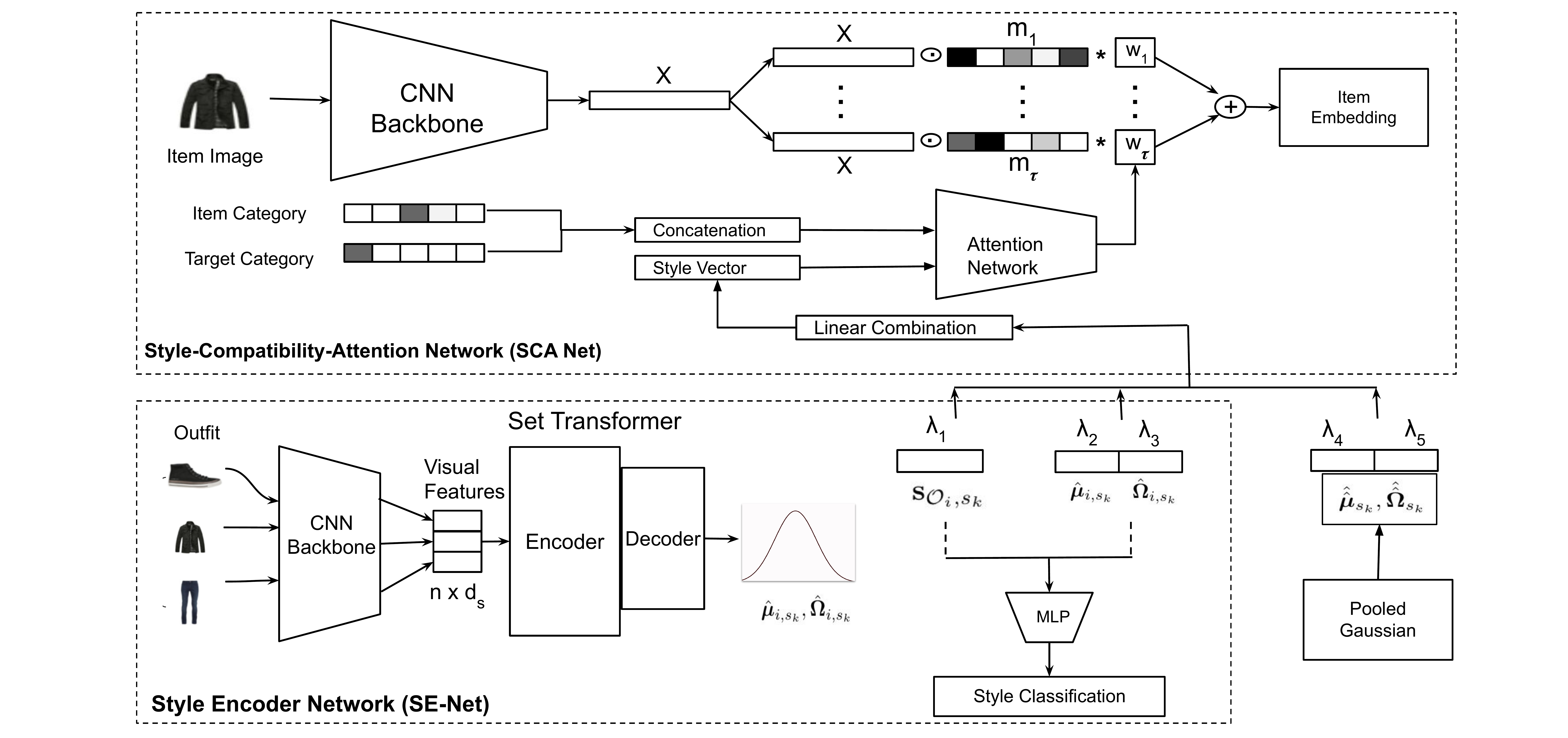}
    \caption{\small{Architecture of  SATCORec. The lower module combines the SE-Net and Style classifier and is trained separately. Item images of an outfit are fed to a CNN to extract visual features which are subsequently passed onto a Set transformer to output a Gaussian distribution. The style classifier is trained using either a random sample or the parameters of the Gaussian. A linear combination of these two along with the parameters of style-specific pooled Gaussian is passed as a feature in the SCA Net module which learns compatibility via attention.}}
    \label{fig:proposed}
    \vspace{-2mm}
\end{figure}

We start with proposing a novel \emph{Style Encoder Network} (SE-Net) which learns a parametric probability distribution representing outfit style using the set transformer \cite{Lee:2019:SetTransformer}, followed by a style classification task further downstream. We extend the compatibility framework of Lin et al.\ \cite{Lin:2020:AmazonFOCIR} to allocate differential importance to features extracted from the image of an item not just based on category information but also on the outfit style, thus complementing SE-Net. We have further modified the compatibility loss in \cite{Lin:2020:AmazonFOCIR} to incorporate style. The entire architecture is shown in Fig. \ref{fig:proposed}. Details of SE-Net and the Style Classifier are provided in Sections~\ref{sec:senet} and \ref{sec:StyleClassifier} respectively. SCA Net and the modified compatibility loss are explained in Section~\ref{sec:scaNet}. We explain the generation of outfits based on individual or mixture of style in Section~\ref{sec:outfit_gen}.

To introduce the notations, let us assume that $m$ explicit styles, say $\mathcal{S} \equiv \{s_1, s_2, \ldots, s_m\}$, are defined in an online portal recommending complete outfits for a user. For an outfit $\mathcal{O}_i$ belonging to style $s_k$ (say $\mathcal{O}_i \lvert s_k$), we assume one of the items within the outfit to be the \emph{anchor item} and the rest is defined as \emph{query set}. We call this \texttt{<anchor item, query set>} as a \emph{positive example} of compatibility. A \emph{negative instance} is one where the anchor item is changed so that it no longer stays compatible with the query set.

\vspace{-4mm}
\subsection{Style Encoder Network}\label{sec:senet}
The process of encoding style of an outfit starts with acknowledging the fact that denoting an outfit as an ordered sequence of items, as is done in some recent work \cite{Han:2017aa,Nakamura:2018:OutfitGenSTyleExtractionBiLSTMAE}, can be seen to be unrealistic. In this paper, we portray an outfit as a set of items which serves two important properties,
    (i) items within an outfit can be termed as permutation invariant, and
    (ii) an outfit is allowed to be of varying length.
This characterization makes the \emph{set transformer} approach \cite{Lee:2019:SetTransformer} an appropriate candidate for our style encoder job. This approach consists of an encoder and a decoder, both of which rely on attention mechanisms to produce a representative output vector.

The idea of representing an individual outfit style by a specific embedding is apt for compatibility training but found to be lacking in the generation context. Since outfit generation is hinged on a single parent item, a pre-defined template and a style, we may not be able to pass any reference outfit to the style encoder. To circumvent this problem, we make the assumption that each $\mathcal{O}_i \lvert s_k$, is generated from some parametric continuous probability distribution thus representing a latent style space. In this paper, we assume that this distribution is Gaussian, although we acknowledge that it can be any other continuous distribution. The parameters of this Gaussian distribution is estimated by the set transformer. In this framework, as can be seen in Figure~\ref{fig:proposed}, the images of an outfit are passed through a pre-trained ResNet18 \cite{He:2016aa} and the corresponding visual feature vectors ($\in {\rm I\!R}^{d_s}$) are fed into the set transformer to provide estimates for the mean vector and co-variance matrix (we assume this to be diagonal). To summarise, the set transformer produces an unique Gaussian distribution for each outfit $\mathcal{O}_i \lvert s_k$,
\[ \displaystyle \mathcal{O}_i | s_k \sim \mathcal{N}(\bm{\mu}_{i, s_k}, \bm{\Omega}_{i, s_k}), \, \text{where } \bm{\Omega}_{i, s_k} = \text{diag}(\sigma_{il, s_k}^2),\quad l=1,\ldots, d_s \text{ and } \mu \in {\rm I\!R}^{d_s}.\]

\begin{figure}[h]
        \includegraphics[scale=0.16]{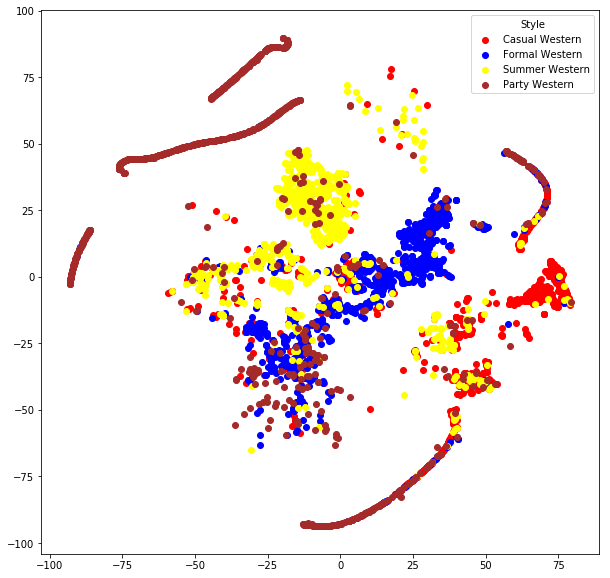}
        \includegraphics[scale=0.16]{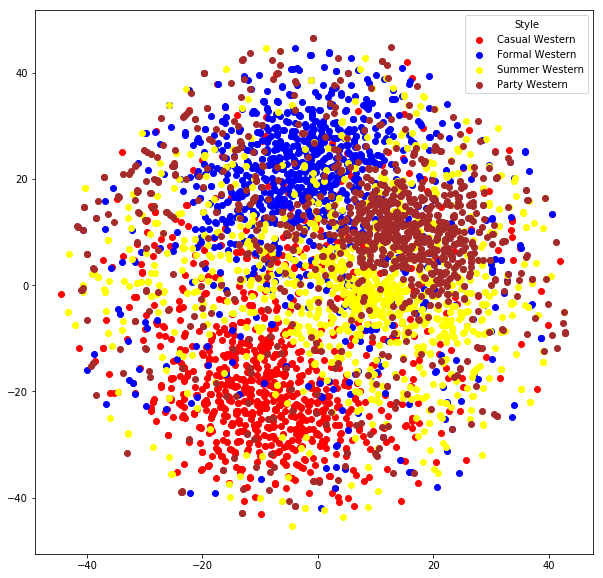}
\caption{\small{t-SNE plots of the sample vectors ($\mathbf{s}_{\mathcal{O}_i, s_k}$) for 4 styles (\textcolor{red}{Casual}, \textcolor{blue}{Formal}, \textcolor{yellow}{Summer}, \textcolor{brown}{Party}). The plot on the left is when these vectors are generated without the KL-divergence loss. Existence of a smooth yet identifiable style latent space is evident in the plot on the right when we introduce the loss. Best viewed in colour.}}
\label{fig:kl_diverg_tsne}
\end{figure}

Here, we additionally impose the restriction that the inferred Gaussian distributions are close to the unit Normal $\mathcal{N}(0, \mathbb{1})$, so that the learnt style space is smooth across the various styles. We achieve this via the KL divergence loss defined in equation~\ref{eqn:style_encoder_loss}. 
\begin{equation}
    \mathcal{L}_{Style} = \text{KL}(\mathcal{N}(\hat{\bm{\mu}}_{i, s_k}, \hat{\bm{\Omega}}_{i, s_k})\, \lvert \lvert \, \mathcal{N}(0, \mathbb{1})) \label{eqn:style_encoder_loss}
\end{equation}
Figure~\ref{fig:kl_diverg_tsne} demonstrates a t-SNE visualisation of random samples drawn from outfit specific Gaussians for 4 different styles. A common and smooth representation space is formed after introducing the KL-loss even though clusters are maintained.  A smooth space is necessary particularly in the generation of outfits with style mixing, as we will see later.

\vspace{-4mm}
The output emanating from the set transformer is passed on for a style classifier job. Depending on the specific variation, an outfit $\mathcal{O}_i \lvert s_k$, we pass either the parameters of the Gaussian ($\theta_{i, s_k} \equiv [\hat{\bm{\mu}}_{i, s_k}, \hat{\bm{\Omega}}_{i, s_k}]$) or a random sample from the Gaussian $\mathbf{s}_{\mathcal{O}_i, s_k} \sim \mathcal{N}(\hat{\bm{\mu}}_{i, s_k}, \hat{\bm{\Omega}}_{i, s_k})$ to the style classifier. We have elaborated on the exact process in section~\ref{sec:StyleClassifier}.

\vspace{-4mm}
\subsection{Style classifier}\label{sec:StyleClassifier}
The SE-Net output vector is passed as a feature to an MLP used to classify the style of the outfit. This supervision ensures that SE-Net captures specific and correct information about the outfit style. The style classification module solves an $m$-class classification problem using an MLP with $N$ layers. The classification loss is thus,

\begin{equation}
    \mathcal{L}_{\text{classif}} = -\sum_{i=1}^{m}y_{s_k} log(\hat{p}(O_i\mid s_k)) \label{eqn:style-classification-loss}
\end{equation}
where $y_{s_k} = 1 \text{, if outfit } O_i$ has style $s_k$ and $\hat{p}(O_i\mid s_k) = \text{MLP}(\mathbf{s}_{\mathcal{O}_i, s_k} \text{ or } \theta_{i, s_k})$. 
The SE-Net and style classifier are trained jointly as a separate module. Post training, we extract a vector ($\mathbf{r}_{\mathcal{O}_i, s_k}$) from this module as style representation of the outfit $\mathcal{O}_i \lvert s_k$ to be passed as a feature to SCA Net. Further, a \emph{global style} representation for a style $s_k$ is given by a pooled Gaussian distribution, aggregating over the parameters of all outfits belonging to that style: $\hat{\hat{\bm{\mu}}}_{s_k} = \frac{1}{n_{s_k}}\sum_{i=1}^{n_{s_k}}\hat{\mu}_{i, s_k}$ and $\hat{\hat{\bm{\Omega}}}_{s_k} =\text{diag}(\hat{\hat{\sigma_{l}^2}})$ where $\hat{\hat{\sigma}}_{l}^2 = \frac{1}{n_{s_k}^2}\sum_{i=1}^{n_{s_k}}\hat{\sigma}_{il}^2$. These global distribution parameters will be used again in the outfit generation step. Equation~\eqref{eqn:style_representation} shows a generic form of style representation vector,
\begin{equation}
    \mathbf{r}_{\mathcal{O}_i, s_k} \equiv \left[\lambda_1\, \mathbf{s}_{\mathcal{O}_i, s_k} +  \lambda_2\,\hat{\bm{\mu}}_{i, s_k} + \lambda_4\,\hat{\hat{\bm{\mu}}}_{s_k}, \lambda_3\hat{\bm{\Omega}}_{i, s_k} + \lambda_5\hat{\hat{\bm{\Omega}}}_{s_k}\right]. \label{eqn:style_representation}
\end{equation}
SATCORec variations, defined in Table~\ref{tab:SATCORec_variations}, are created by setting values for each $\lambda_j$. Also note that, we pass $\mathbf{s}_{\mathcal{O}_i, s_k}$ to the style classifier for \emph{SATCORec-r}, \emph{SATCORec-($p_m$+$g_m$)} and \emph{SATCORec-(r+$g_m$)} and $\theta_{i, s_k}$ for the rest. It is possible to set $\lambda$ as unknown and learn it.


\vspace{-4mm}

\begin{table}[h]
\caption{\small{Variations of SATCORec that have been experimented with.}} \label{tab:SATCORec_variations}
\footnotesize
\begin{tabular}{|l|c|c|c|c|c|l|c|c|c|c|c|}
\hline
 & $\lambda_1$ & $\lambda_2$ & $\lambda_3$ & $\lambda_4$ & $\lambda_5$ &
 & $\lambda_1$ & $\lambda_2$ & $\lambda_3$ & $\lambda_4$ & $\lambda_5$ \\ \hline
SATCORec-r & 1 & 0 & 0 & 0 & 0 & 
SATCORec-p & 0 & 1 & 1 & 0 & 0 \\ \hline
SATCORec-($p_m$+$g_m$) & 0 & $\lambda$ & 0 & 1 & 0 & 
SATCORec-(p+g) & 0 & $\lambda$ & $\lambda$ & 1 & 1 \\ \hline
SATCORec-(r+$g_m$) & $\lambda$ & 0 & 0 & 1 
& 0 & \multicolumn{6}{c|}{} 
\\ \hline
\end{tabular}
\end{table}

\vspace{-6mm}
\subsection{SCA Net}\label{sec:scaNet}
 
We have extended the CSA-Net framework developed by Lin et al.\ in \cite{Lin:2020:AmazonFOCIR} to incorporate the concept of style while learning item-item compatibility. In \cite{Lin:2020:AmazonFOCIR}, the image of an anchor item ($I^a$) within an outfit is passed through a ResNet18, which acts as the CNN backbone. The embedding output vector ($\mathbf{x}$) of size 64 is multiplied by $k$ learnt masks ($\mathbf{m}_1, \ldots, \mathbf{m}_\tau$) that help to learn the subspaces. The anchor item category ($c^a$) and a query set item (referred to as \emph{target}) category ($c^t$) information are consumed as 1-hot  encoded vectors to estimate a set of subspace attention weights ($\omega_1, \ldots, \omega_\tau$). A weighted average of the masked embeddings results in the final embedding of the anchor item.

We simply extend the CSA-Net algorithm by providing the style representation ($\mathbf{r}_{\mathcal{O}_i, s_k}$) from SE-Net as an additional input in the estimation of attention weights. Thus, we define the final embedding as,
\[ \displaystyle f_{\mathcal{O}_i, a}^{s_k} = \psi(I^a, c^a, c^t, \mathbf{r}_{\mathcal{O}_i, s_k}) = \sum_{j=1}^\tau (\mathbf{x} \odot \mathbf{m}_j) \times \omega_{j, \mathbf{r}_{\mathcal{O}_i, s_k}}.\]
Here, $\psi(\cdot)$ represents the SCA network.

The SCA net uses the triplet loss for learning compatibility, similar to some current methods \cite{Tan:2019:LSCWES,Vasileva:2018:LTAE}. We represent the average distance between a positive item and remaining items in the outfit as $D_p^{s_k}$, same as CSA-Net. The multiple distances corresponding to the negatives are aggregated  as $D_N^{s_k}$. The overall compatibility loss conditional on style is thus defined as,
\begin{equation}
    \mathcal{L}_{compat} = \max(0, D_p^{s_k} - D_N^{s_k} + m), \label{eqn:compatibility_loss_1}
\end{equation}

We introduce one more loss function to account for penalisation when the wrong style is specified for an outfit. Given $\mathcal{O}_i | s_k$, we pass the style representation vector corresponding to a different style $s_q$, and compute the same distance metrics as above, and use them in the following loss function:
\begin{equation}
    \mathcal{L}_{stylecompat} = \max(0, D_p^{s_k} - D_p^{s_q} + m). \label{eqn:compatibility_loss_2}
\end{equation}
The overall loss is defined as the weighted sum of these four individual losses:
\begin{equation*}
    \mathcal{L}_{overall} = \sum_p \alpha_p\, \mathcal{L}_p, \quad p \in \{\text{KL, classification, compatibility, style-compatibiliy}\} \label{eqn:total_loss}
\end{equation*}

\vspace{-6mm}
\subsection{Outfit generation}\label{sec:outfit_gen}
A globally optimal outfit generation task is non-trivial since it is infeasible to look into all possible combinations. An approximate solution based on the well known \emph{beam search method} \cite{Zhang:2020aa} is provided in this case. Note that to create an outfit for a user based on a chosen parent item, a given template and a specific style, we need a style representation vector to rank compatible items. If there is a reference outfit present, then this job is trivial. In the alternative case, we assume the pooled parameters to be representative of style for all the variations within SE-Net. To generate an outfit based on mixing of styles, we simply pass a linear combination of style representation vectors ($\alpha \mathbf{r}_{\mathcal{O}_i, s_k} + \beta \mathbf{r}_{\mathcal{O}_i, s_l}$) and rank compatible items.

\section{Experimental Evaluation}
In this section, we elaborate the dataset, metrics, baselines, implementation details and the different results testing the compatibility as well as style preservation power of the algorithms.   \\

\noindent{\bf Dataset creation and metrics:}
We have annotated $\sim$100K outfits in two stages. At {\bf first}, we worked with fashion experts to get approximately 5000 outfits curated with 8 style annotations, namely [\emph{party}, \emph{outdoor}, \emph{summer}, \emph{formal}, \emph{athleisure}, \emph{winter}, \emph{causal}, \emph{celeb}]. Each annotated outfit consists of the images of individual items and and its style. There 6 high level item categories, [\texttt{top-wear},  \texttt{bottom-wear}, \texttt{foot-wear},\texttt{accessory}, \texttt{clothing-accessory}, \texttt{wholebody}]. In the {\bf second} stage, we augmented the outfit set using a simple attribute based similarity algorithm, where we used attributes like brand, colour, pattern, sleeve, etc. to get top-k similar products for an item in an outfit. Given an outfit, we removed one item from the original outfit and gave approx. top-10 similar candidates as options for replacement to human taggers for verification of compatibility and style of the new outfit. We repeated this for all item in an outfit and for all outfits in the outfit set. This operation expanded the data to $\sim$100K outfits, which are then divided into train, test and validation splits in 70:20:10 ratio. The overall frequency for each style type is given in Table~\ref{tab:outfit_stats}.

Fill-in-the-blank (FITB) \cite{Vasileva:2018:LTAE} and Compatibility AU-ROC are well known metrics used to evaluate an outfit compatibility model \cite{Han:2017aa,McAuley:2015:StylesSubstitutes}. Both these approaches involve creating negative items corresponding to each item of an outfit. To test performance at various levels of difficulty, we generate two types of negative items, \emph{soft negatives} where negative sampling is done from existing categories; and \emph{hard negatives} where we sample negatives from more fine-grained categories such as tops, t-shirts, heels etc. For each outfit, 5 replications for negative sampling are done and the mean metric values are reported. Note that the fine-grained category information is not used for training.
\vspace{-4mm}

\begin{table}[h]
\caption{\small{Distribution of curated outfits across different styles}} \label{tab:outfit_stats}
\begin{adjustbox}{max width=\textwidth}
\begin{tabular}{|l|l|l|l|l|l|l|l|l|l|}
\hline
                    & \textbf{Party} & \textbf{Outdoor} & \textbf{Summer} & \textbf{Formal} & \textbf{Athleisure} & \textbf{Winter} & \textbf{Casual} & \textbf{Celebrity} & \textbf{Total} \\ \hline
\# of Train Outfits & 8183  & 6280    & 7061   & 5136   & 16232      & 16028  & 5194   & 5424      & 69538 \\ \hline
\# of Valid Outfits & 1174  & 1001    & 1204   & 840    & 1981       & 2135   & 791    & 808       & 9934  \\ \hline
\# of Test Outfits  & 3018  & 1937    & 2551   & 1648   & 2506       & 4695   & 2034   & 1480      & 19869 \\ \hline
\end{tabular}
\end{adjustbox}
\end{table}
\vspace{-4mm}

\noindent{\bf Implementation details:}
We used ResNet18 as the CNN backbone to extract visual features in both the modules of SATCORec. We do not train the entire ResNet18 but instead only the last convolutional block and an additional fully connected layer. Embeddings are of size 64 as is conventional in other state-of-the-art compatibility learning methods \cite{Lin:2020:AmazonFOCIR,Vasileva:2018:LTAE}.

Inside SE-Net, we use the SAB Set Transformer \cite{Lee:2019:SetTransformer} with hidden dimension $d_z = 32$ and 2 heads. We use 2 fully connected MLP layers for classification. An Adam optimizer \cite{adam} with mini batches of 128 outfits and a learning rate of $5 \times 10^{-5}$ is used. Note that, we have trained and frozen the SE-Net module separately. We used the Adam optimizer again to train SCA Net with a mini-batch size of 32 triplets, learning rate of $1 \times 10^{-5}$ and 5 subspaces. The \emph{Attention network} first transforms the concatenated one-hot-encoded category and the style representations to 32 dimensions each using a single fully connected layer and then concats the two to pass it to 2 fully connected layers which output the 5 subspace attention weights. The margin within the triplet loss was set to 0.3 and the weights for $\mathcal{L}_{compat}, \mathcal{L}_{stylecompat}$ and $\mathcal{L}_{Style}$ were set to $1, 0.5$ and $0.05$ respectively.

\noindent{\bf Baselines: }
We compare the performance of SATCORec against that of state-of-the-art techniques on the basis of the multiple metrics to demonstrate its efficacy in style conditional outfit generation and compatibility learning. Note that we use the same CNN backbone and embedding size for all the baselines. Additionally, the same 6 categories have been used for all the methods, even for those requiring fine-grained category information. The following are used as baselines (a). \textbf{CSA-Net \cite{Lin:2020:AmazonFOCIR}}, (b). \textbf{Type Aware \cite{Vasileva:2018:LTAE}}, (c). \textbf{TransNFCM \cite{yang2019transnfcm}}, (d). \textbf{Theme Matters \cite{Lai:2020:ThemeMatters}}, (e). \textbf{BPR-DAE \cite{Song:2017:BPR-DAE}}. 
For each of the methods we follow the same architecture parameters which the paper specifies. Except {\bf Type aware}, whose code was available, we have implemented all of the baselines from scratch. For {\bf Theme Matters} we have first taken the type aware code and built upon it as is defined in the paper. In \textbf{BPR-DAE}, the method is specified for only for 2 categories, and we extend it for outfits with multiple items.
\vspace{-2mm}

\begin{table}[h]
\caption{\small{Comparison of compatibility learning for the baselines and SATCORec variations. We compute FITB and compatibility AU-ROC with hard and soft negatives separately. The style entropy for each methods are also tabulated. Using parameters or random sample from outfit style specific Gaussian is clearly the leader with respect to compatibility measures.}}
\label{tab:fitb-compatauc-metrics}
\begin{adjustbox}{max width=\textwidth}
\begin{tabular}{|l|c|c|c|c|c|c|}
\hline
\multirow{2}{*}{\textbf{Method}} & \multicolumn{2}{c|}{\textbf{FITB}} & \multicolumn{2}{c|}{\textbf{Compat. AU-ROC}} & \multirow{2}{*}{\textbf{Entropy}} \\
\cline{2-5}
 & \textbf{HN }       & \textbf{SN }       & \textbf{HN} & \textbf{SN} & \\
\hline
TypeAware                 & $30.7 \pm 0.17$ & $34.85 \pm 0.25$  & $52.62 \pm 0.06$  & $55.51 \pm 0.21$ & 0.49\\ \hline
BPR-DAE & $31.16 \pm 0.15$ & $31.21 \pm 0.12$ & $55.83 \pm 0.09$ & $55.76 \pm 0.08$ & 0.43\\ \hline
TransNFCM                 & $31.53 \pm 0.17$ & $36.47 \pm 0.33$  & $51.84 \pm 0.07$  & $57.78 \pm 0.08$ & 0.50\\ \hline
Theme Matters                 & $38.53 \pm 0.17$ & $63.2 \pm 0.21$  & $85.4 \pm 0.15$  & $93.85 \pm 0.1$ & 0.61\\ \hline
CSA-Net  & $53.14 \pm 0.17$ & $67.05 \pm 0.25$ & $94.42 \pm 0.03$ & $96.3 \pm 0.03$ & 0.48 \\ \hline
SATCORec-r                & ${\bf 53.32} \pm 0.18$ & $66.63 \pm 0.15$ & $94.47 \pm 0.02$ & $95.99 \pm 0.04$ & ${\bf 1.09}$ \\ \hline
SATCORec-p & $52.06\pm 0.10$ & ${\bf 67.31} \pm 0.14$ & ${\bf 94.78} \pm 0.02$ &	${\bf 96.47}\pm 0.02$ & $0.97$ \\ \hline
SATCORec-(p+g) & $46.56\pm 0.05$ & $61.03 \pm 0.17$ & $88.41\pm 0.02$ & $90.10 \pm 0.02$ & $0.78$ \\ \hline
SATCORec-(r+$g_m$) & $47.61 \pm 0.12$ & $60.70\pm 0.06$ & $88.88\pm 0.06$ & $91.34\pm0.02$ & $0.12$ \\ \hline
SATCORec-($p_m$+$g_m$) &  $49.73\pm0.05$& $63.02\pm 0.11$& $90.96\pm 0.05$& $92.25 \pm 0.02$ & $0.63$\\ 
\hline
\end{tabular}
\end{adjustbox}
\end{table}
\vspace{-8mm}

\subsection{Compatibility Experiment:} \label{sec:comp-exp}
FITB and compatibility AU-ROC are computed separately on the hard and soft negative datasets for variations of SATCORec and the baselines and presented in table~\ref{tab:fitb-compatauc-metrics}. A preliminary sweep of the results clearly differentiates the performance of Theme Matters, CSA-Net and SATCORec variations from the rest. CSA-Net is based on subspace based attention mechanism, which is the state-of-the-art in learning outfit item compatibility, and SATCORec makes use of the same framework. It is surprising that Theme Matters performs better than TypeAware since both have the same compatibility learning framework. This performance bump is caused due to these methods incorporating complete outfit loss in their learning \cite{Lin:2020:AmazonFOCIR}.

SATCORec-p is the best performing model in the group, winning in 3 out of 4 cohorts. We think that the outfit-level Gaussian parameters capture sufficient information about the parent style of the outfit as well as variations within. The random sampling of the space can also capture the basic information of a style category, resulting in the healthy performance of SATCORec-r. The other variations do not perform well, probably because of ignoring individual or overall uncertainty. 

\vspace{-4mm}
\subsection{Style Experiments}
\label{sec:style-exp}
Given that our methods show better performance than others in compatibility learning, we now compare their performance vis-a-vis style. We look at two specific style comparison metrics and discuss a characteristic that our method has, but is absent in style-independent methods. Statistical comparisons for our metrics and further qualitative results will be added over time in this link: \url{https://harshm121.github.io/project_pages/satco_rec.html}.

\vspace{-4mm}
\subsubsection{Style Entropy:} 
A user would get maximum utility if her top-wear can be part of outfits belonging to a large number of style categories, i.e.\ the portal is able to recommend from a wide range of styles. Say given an anchor item, we want to recommend a total of $n$ outfits from $k$ styles. SATCORec, using the style-handle, can produce a ranked lists of outfits conditioned on each of the $k$ styles. We choose the top $\floor{n/k}{}$ or  $\ceil{n/k}{}$ outfits from each style specific list. Style independent methods will get its top-$n$ outfits as per the general compatibility rank, thus oblivious to their reference styles. We use the entropy measure on style to compare the final lists. A higher entropy would mean that the compatibility framework is not restrictive to a single or small number of styles. For this, we select the list of all those outfits which have the same anchor item, but belong to different styles. From this list, we pick those instances where SATCORec is able to correctly predict the items of an outfit given a style. We then choose the top outfit from each each style, thus forcing $n=k=6$, and present the result in table~\ref{tab:fitb-compatauc-metrics}, column \emph{Entropy}. Again, SATCORec-r (slighlty better) and SATCORec-p outperform all other methods, implying that they are able to recommend outfits corresponding to most of the styles feasible for the anchor item. On manual inspection we also find that style-independent methods are biased towards the most prevalent style in the training data set. Henceforth, we will consider only the top performing variations, SATCORec-r and SATCORec-p. 

\vspace{-4mm}
\subsubsection{Style-specific selection accuracy and ranking:} SACTORec-r is also seen to be superior in some other metrics we compute like MRR, Avg rank etc. Table~\ref{tab:child-selection-ranking-results} - \emph{Metric} captures this for the three style-dependent methods. Given a method, we have taken each outfit and calculated the compatibility scores conditional on all the available styles. We record the outfit rank corresponding to the style it actually belongs compute the metrics based on them. Fig \ref{fig:overview} presents an example for an anchor top-wear and the style conditional compatibility scores for each of the outfits comprising only of bottom-wear. We see that the scores are highest (top ranked) for the style in which the outfit actually belongs.

\vspace{-4mm}
\subsubsection{Other Metrics:} We make use of the list of outfits used in the calculation of style entropy again to understand the efficacy of the algorithms. Note that this list has outfits from different styles but common anchor item. For each such anchor item, conditional on the style, we check the top-1 accuracy of selecting the right child item in the outfit. To understand \emph{accuracy}, we refer again to Fig \ref{fig:overview}, where accuracy for Bottomwear1 equals 1 since the inferred rank corresponding to actual style is lowest. Table~\ref{tab:child-selection-ranking-results} - \emph{Parent-Child} shows the results for various parent-child category combinations. Here SATCORec-p performs much better, although when we were checking column-wise ranking, it was behind SATCORec-r.

\vspace{-4mm}

\begin{table}[h]
\caption{\small{The upper section of the table contains metrics on outfit ranks conditional on style while the lower section provides the percentage of correct selection of compatible item for anchor items with outfits across various styles.}}
\label{tab:child-selection-ranking-results}
\resizebox{\columnwidth}{!}{
\begin{tabular}{|c|c|c|c|c|}
\hline
 & & \textbf{SATCORec-r} & \textbf{SATCORec-p}& \textbf{Theme Matters}\\ \hline
\multirow{4}{*}{\textbf{Metric}} & MRR of correct style & \textbf{0.8844}  & 0.7676 & 0.6213 \\
\cline{2-5}
& Correct style on 1st rank & \textbf{80.94} & 59.36 & 42.37 \\
\cline{2-5}
& Correct style in top 3 ranks & 95.00 & \textbf{95.10} & 76.51 \\
\cline{2-5}
& Avg rank of the correct style & \textbf{1.4} & 1.7 & 2.5\\ \hhline{|=|=|=|=|=|}
\multirow{4}{*}{\textbf{Parent-Child}} & Topwear - Bottomwear & 66.74 & \textbf{77.33} & 50.32 \\
\cline{2-5}
& Bottomwear - Topwear & 72.02 & \textbf{86.65} & 57.92 \\
\cline{2-5}
& Topwear - Footwear & 65.79 & \textbf{75.97} & 59.73 \\
\cline{2-5}
& Bottomwear - Footwear & 69.81 & \textbf{80.13} & 62.79\\\hline 
\end{tabular}
}
\end{table}

\vspace{-4mm}

\noindent\textbf{Style-Specific fine-grained category selection in outfit generation:} 
For each style, there can be multiple child-items which may match an anchor item, however, a good recommendation system would mostly output the items which differentiate the outfit from other styles. To check this phenomenon, for each style, we  determine the most discriminating child-items \cite{tf-idf}, in terms of fine-grained categories e.g. \emph{skirt} is a fine-grained category in bottomwears which most prominently shapes a casual style. Note that this is different from the most popular item across styles, say for example \emph{jeans}. We posit that a superior algorithm would more frequently output such discriminative categories as a likely-match for a style. Style specific and overall results are shown in Table~\ref{tab:generated_outfit_analysis}, we see in almost all the cases, SATCORec's output chooses discriminative fine-grained categories significantly higher number of times than the other baselines.

\begin{table}[h]
\caption{\small{Comparison of style-specific fine-grained categories chosen by different methods.}}
\label{tab:generated_outfit_analysis}
\begin{adjustbox}{max width=\textwidth}
\begin{tabular}{|l|l|l|l|l|l|l|l|l|l|}
\hline
\textbf{Method} & \textbf{Party} & \textbf{Outdoor} & \textbf{Summer} & \textbf{Formal} & \textbf{Athleisure} & \textbf{Winter} & \textbf{Casual} & \textbf{Celeb} & \textbf{Overall} \\ \hline
TypeAware         & 28.33         & 29.22           & 10.24           & 33.54          & 19.52          & 18.10 & 2.67          & 15.92 & 19.30 \\ \hline
BPR-DAE &  28.19 & 17.07 & 17.74 & 36.26 & 31.64 & 29.05 & 23.42 & 19.05 & 25.64\\
\hline
TransNFCM         & 12.78         & 25.72           & 3.09            & 23.84          & 30.01           & 21.21           & 0.00          & 27.86 & 18.36 \\ \hline
CSA-Net         & 34.63          & 26.79            & 13.98           & 35.44           & 28.69               & 26.94           & 11.00           & 27.11 & 25.38        \\ \hline
Theme Matters   & 34.26          & 24.20            & 7.48            & 24.68           & 14.21               & 30.05           & 18.00           & 9.95           & 21.39            \\ \hline
SATCORec-r       & \textbf{50.56} & \textbf{32.12} & 19.84 & 45.78 & \textbf{38.65} & 39.31 & 18.17 & 25.62          & \textbf{34.27}   \\ \hline
SATCORec-p & 38.59 & 21.89 & \textbf{23.06} & \textbf{47.26} & 37.18 & \textbf{40.92} & \textbf{24.09} & \textbf{28.09} & 32.96\\
\hline
\end{tabular}
\end{adjustbox}
\end{table}

\begin{figure}[ht]
    \centering
    \includegraphics[width=0.75\linewidth]{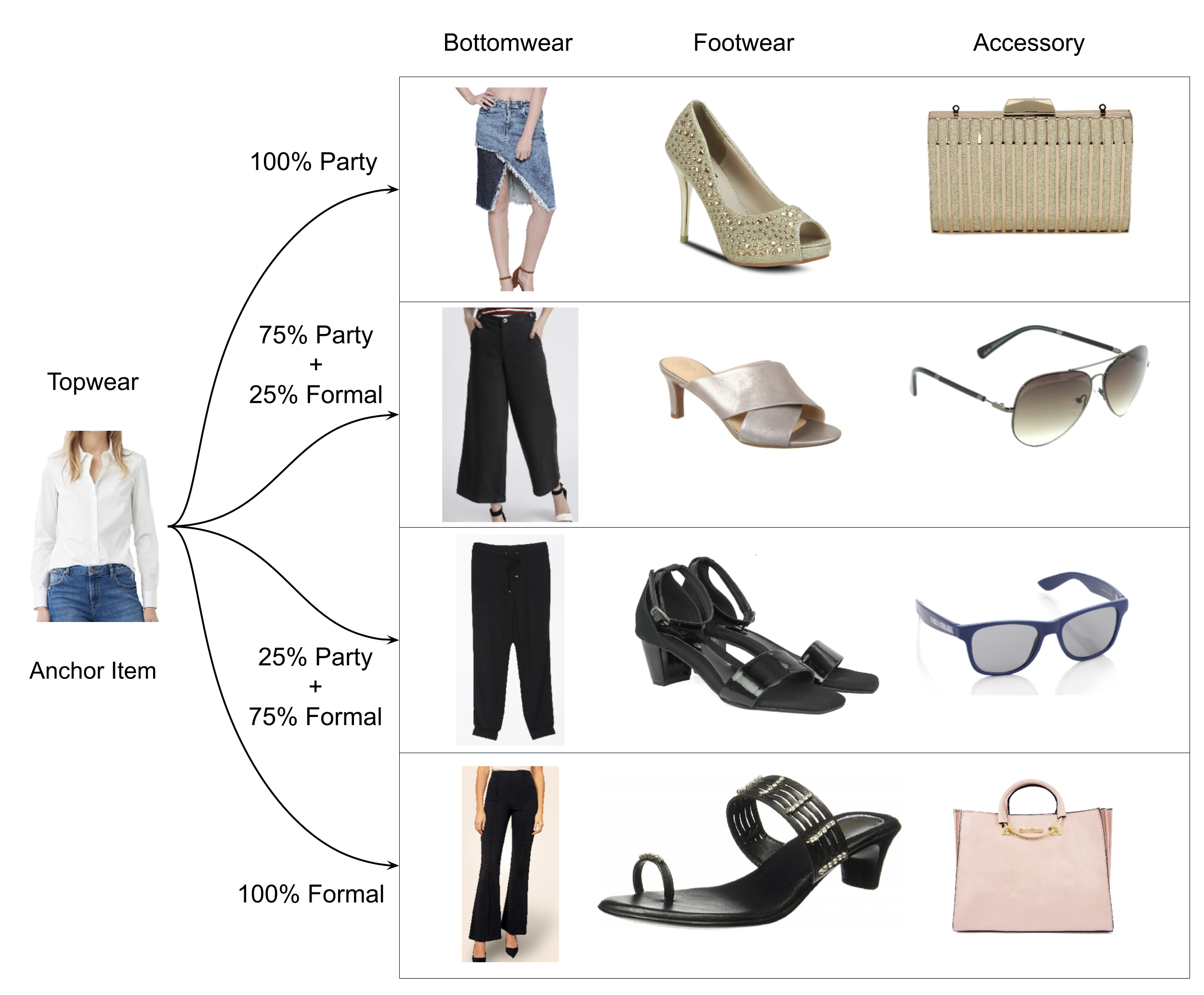}
    \caption{\small{Here we demonstrate the ability of our method to mix styles in outfit generation. Given the anchor item from top-wear, the top and bottom rows correspond to outfits generated from the two very separate styles: \emph{Party} and \emph{Formal}. The outfits in between them are generated by passing a weighted style vector for each of those two styles, thereby creating a nice blend.
    }}
    \label{fig:transition_outfit_example_1}
\end{figure}

\noindent\textbf{Blending of Styles:} 
We have also checked the ability of SATCORec to generate outfits that are a linear combinations of different styles. We observe a smooth blending of the styles, also a higher (lesser) weight of a particular style (in the linear combination) results  in the presence of more (less) items resembling that style in the generated outfits (Figure \ref{fig:transition_outfit_example_1}). 
We will provide an web-based app along with the final version of the paper if accepted where a user would be able to explore different such combinations. 
\vspace{-4mm}

\vspace{-10mm}
\section{Conclusion}
The novelty of the paper lies in developing a  Style-Attention-based Compatible Outfit recommendation and generation framework, SATCORec, 
utilizing high-level categories. SATCORec employs a Style-Compatibility-Attention Network~-~SCA Net and a Style Encoder Network~-~SE-Net. The SE-Net uses the Set Transformer to extract outfit style features, which is used to  provide style-specific sub-space attention to individual items. The extensive style experiments establish the power of SATCORec in recommending with high accuracy a broader collection of compatible outfits across different styles to users. 
More interestingly, SATCORec chooses items which can make a pronounced style statement. 
Since in this paper we have focused on compatibility and employed a traditional beam search for outfit generation, an immediate future work would be to explore more sophisticated generation algorithms. 

\bibliographystyle{splncs04}
\bibliography{ref}
%




\end{document}